\documentclass[english, 12pt, letter]{article}
\usepackage{amsmath,amssymb,amsthm}
\usepackage{natbib}
\usepackage[nodisplayskipstretch]{setspace}
\usepackage{bm}
\usepackage{threeparttable}
\usepackage{graphicx}
\usepackage{enumerate}
\usepackage{commath}			
\usepackage{bibentry}
\usepackage{booktabs}
\usepackage[titletoc,title]{appendix}
\usepackage{multirow}
\usepackage{arydshln}
\usepackage{tabularx}
\usepackage{dsfont}
\usepackage[font={footnotesize}]{subcaption}
\usepackage[left=1in, right=1in, top=1in, bottom=1in]{geometry}
\usepackage{pifont}
\newcommand{\xmark}{\ding{55}}%

\usepackage[colorinlistoftodos,textsize=tiny]{todonotes}
\allowdisplaybreaks

\usepackage[
	breaklinks,
	colorlinks=true,
	pagebackref=false,
	pdfauthor={},%
	pdftitle={},%
	citecolor=blue,
	linkcolor=blue,
	urlcolor=blue
	]{hyperref}      

\newtheorem{thm}{Theorem}
\newtheorem{lem}[thm]{Lemma}
\newtheorem{assumption}[thm]{Assumption}
\newtheorem{prop}[thm]{Proposition}

\newtheoremstyle{assumption}{1ex}{1ex}%
      {\it}
      {}
      {\bf}
      {.}
      {5pt}
      {\thmname{#1}\thmnumber{ #2}*\thmnote{ \slshape{(#3)}}} 
\theoremstyle{assumption}	

\newtheoremstyle{remark2}{1ex}{1ex}%
      {}
      {}
      {\bf}
      {.}
      {5pt}
      {\thmname{#1}\thmnumber{ #2}\thmnote{ \slshape{(#3)}}} 
\theoremstyle{remark2}
\newtheorem{rem}[thm]{Remark}
\newtheorem{defn}[thm]{Definition}
\newtheorem{example}[thm]{Example}

\usepackage{babel}
\input{ee.sty}

\makeatletter
\renewenvironment{proof}[1][\bfseries\proofname]{\par
   \pushQED{\qed}%
   \normalfont \topsep6\p@\@plus6\p@\relax
   \trivlist
   \item[\hskip\labelsep
     #1\@addpunct{:}]\ignorespaces
}{%
   \popQED\endtrivlist\@endpefalse
}

\providecommand{\leftsquigarrow}{%
  \mathrel{\mathpalette\reflect@squig\relax}%
}
\newcommand{\reflect@squig}[2]{%
  \reflectbox{$\m@th#1\rightsquigarrow$}%
}
\makeatother

\newcommand{\Comments}{1}
\newcommand{\mynote}[2]{\ifnum\Comments=1\textcolor{#1}{#2}\fi}
\newcommand{\mytodo}[2]{\ifnum\Comments=1%
  \todo[linecolor=#1!80!black,backgroundcolor=#1,bordercolor=#1!80!black]{#2}\fi}

\ifnum\Comments=1               
\fi

\ifnum\Comments=1               
\fi

\newcommand{\marg}{\operatorname{marg}}
\newcommand{\cop}{\operatorname{cop}}
\newcommand{\comp}{\operatorname{comp}}
\newcommand{\CRPS}{\operatorname{CRPS}}
\newcommand{\p}{\operatorname{P}}

\newcommand{\R}{\mathbb{R}}
\newcommand{\A}{\mathsf{A}}

\newcommand{\F}{\mathcal{F}}
\newcommand{\C}{\mathcal{C}}

\newcommand{\T}{T}
\newcommand{\Cop}{T_{\cop}}

\newcommand{\lorder}{\preceq}
\newcommand{\slorder}{\prec}

\newcommand{\lex}{\lorder_{\mathrm{lex}}}

\newcommand{\slex}{\slorder_{\mathrm{lex}}}

\begin{document}

\baselineskip18pt
\renewcommand\floatpagefraction{.9}
\renewcommand\topfraction{.9}
\renewcommand\bottomfraction{.9}
\renewcommand\textfraction{.1}
\setcounter{totalnumber}{50}
\setcounter{topnumber}{50}
\setcounter{bottomnumber}{50}
\abovedisplayskip1.5ex plus1ex minus1ex
\belowdisplayskip1.5ex plus1ex minus1ex
\abovedisplayshortskip1.5ex plus1ex minus1ex
\belowdisplayshortskip1.5ex plus1ex minus1ex

\title{\textbf{How to Compare Copula Forecasts?}
}

\author{
	Tobias Fissler\thanks{RiskLab, ETH Zurich, R\"{a}mistrasse 101, CH--8092 Zurich, Switzerland, \href{mailto:tobias.fissler@math.ethz.ch}{tobias.fissler@math.ethz.ch}.}
\and 
	Yannick Hoga\thanks{Faculty of Economics and Business Administration, University of Duisburg-Essen, Universit\"atsstra\ss e 12, D--45117 Essen, Germany, \href{mailto:yannick.hoga@vwl.uni-due.de}{yannick.hoga@vwl.uni-due.de}.}
}

\date{\today}
\maketitle

\begin{abstract}
\noindent This paper lays out a principled approach to compare copula forecasts via strictly consistent scores.
We first establish the negative result that, in general, copulas fail to be elicitable, implying that copula predictions cannot sensibly be compared on their own.
A notable exception is on Fr\'echet classes, that is, when the marginal distribution structure is given and fixed, in which case we give suitable scores for the copula forecast comparison.
As a remedy for the general non-elicitability of copulas, we establish novel multi-objective scores for copula forecast along with marginal forecasts.
They give rise to two-step tests of equal or superior predictive ability which admit attribution of the forecast ranking to the accuracy of the copulas or the marginals.
Simulations show that our two-step tests work well in terms of size and power.
We illustrate our new methodology via an empirical example using copula forecasts for international stock market indices. \\


\end{abstract}


\section{Motivation}\label{Motivation}


Modeling the dependence between several outcomes is of interest in forecasting economic, financial or meteorological variables, among others.
Since the seminal work of \citet{Skl59}, 
it is known that copulas provide a complete description of the dependence structure of a random vector with continuous marginal distributions.
Therefore, applications of copula modeling are by now widespread in forecasting contexts. For instance, in finance, copulas have been used for forecasting the Value-at-Risk \citep{GHS09}, for portfolio construction \citep{Pat04}, for predictive modeling of exchange rates \citep{CKL13}, and for systemic risk prediction \citep{BC19}.
We refer to the review articles by \citet{MR12} and \citet{Pat13} for many more examples, and to \citet{MLT13} for a meteorological forecasting application. 

A key step in copula modeling is the choice of a suitable parametric family, such as the Gaussian copula, the $t$-copula, an Archimedean copula, or any mixture thereof.
If dependence varies over time, an additional choice involves the driving process for the copula parameters, such as the dynamics proposed by \citet{Pat06} or the more recent generalized autoregressive score (GAS) dynamics of \citet{CKL13}.
The sheer number of choices that a practitioner is faced with in copula modeling calls for statistically sound tools for model selection and model comparison.

The main purpose of this paper is to develop such tools.
Model comparison commonly hinges on strictly consistent scoring functions (often also called loss functions).
Taking as arguments the forecast and the observation, these scoring functions are strictly minimized in expectation by the true model choice.
If such a strictly consistent score exists for a certain functional of interest, such as the mean or a quantile, this functional is called \emph{elicitable}.
E.g., the mean functional is elicitable, with the square loss as a strictly consistent scoring function.
But not all functionals are elicitable, e.g., variance and Expected Shortfall fail to be elicitable \citep{Gne11}. 
The reason is that these functionals do not have convex level sets (CxLS), which is a necessary condition for elicitability \citep{Osband1985, Gne11}.
Our first main result, Proposition~\ref{prop:CxLS of cop}, shows that copulas generally fail to be elicitable, due to a lack of the CxLS property.

A notable exception to our non-elicitability result is on Fr\'echet classes, that is, on classes $\F$ of multivariate distributions where the distributions only differ in terms of their \emph{dependence} structure, i.e., their copulas, but their \emph{marginal distribution} structure is the same, i.e., for any  $F, G\in\F$, it holds that $F_i = G_i$ for all marginal cumulative distribution functions $F_i$ and $G_i$ ($i=1, \ldots, d$).
In practical terms, this means that the \emph{correct marginal model} is known, and the forecaster is only concerned with model choice for the copula.
We provide possible strictly consistent scores for this important situation.
In the previous literature on copula choice, researchers have often relied on the Kullback--Leibler information criterion (KLIC) to decide between different copula models with fixed marginals \citep{CF06,DPV10}.
Our positive result not only formally justifies this practice by grounding it in a sound decision-theoretic framework, but it also points towards a wealth of other scores that can be used (see Proposition~\ref{prop:elicitability for fixed marginals}).

To mitigate the problem of (in general) non-elicitable copulas, it is possible to adopt a strategy pioneered by \citet{Osband1985}.
He suggests to turn a non-elicitable functional into an elicitable one by pairing it with an appropriate auxiliary functional; the pairs (variance, mean) and (Value-at-Risk, Expected Shortfall) being prime examples of this strategy \citep{FZ16a}.
Applied to our situation, we can pair the copula with the marginal distributions.
This amounts to specifying the full joint distribution, which is elicitable with proper scores readily available, e.g., the log-score or the continuous ranked probability score \citep{GR07}.
However, when comparing the full joint distribution with proper scores, we cannot determine if differences in the ranking are due to superior marginal forecasts \textit{or} due to superior copula forecasts.

We address this \emph{attribution problem} by drawing on the recently proposed notion of multi-objective elicitability \citep{FH24}.
This concept works with \textit{bivariate} (instead of real-valued) scores, which are uniquely minimized in expectation by the true forecast with respect to the lexicographic order.
Such scores are called \textit{strictly consistent multi-objective scores}.
Our second main contribution is to show that strictly consistent multi-objective scores exist for the tuple consisting of the copula and all marginal distributions (see Theorem~\ref{thm:mo copula}).

The third main contribution of this article is to show how the multi-objective scores from Theorem~\ref{thm:mo copula} can be used in two-step tests for predictive ability.
In the first step, the marginals are tested for equal predictive accuracy.
The second step then tests for equal or superior predictive ability in the copula forecasts.
This procedure allows a rejection of the null to be clearly attributed to the accuracy of the marginal or of the copula forecast (see Section~\ref{Some Two-Step Tests}).
At each step, our tests are akin to standard Diebold--Mariano tests \citep{DM95}.
However, critical values are computed in such a way that size is controlled across the two steps.

Monte Carlo simulations demonstrate that our two-step tests hold size even in relatively small samples (Section~\ref{Simulations}).
Also the power of the tests in discriminating between copula forecasts of different quality is high.
Our study also reveals that the tests' power in the copula component is higher when the marginal forecasts are less seriously misspecified.
Moreover, we find that attribution also works well in the sense that differences in predictive ability of the marginals (copula) is detected almost exclusively in the first step (second step).

We illustrate the use of our two-step tests in an application to copula predictions for returns from international stock market indices (Section~\ref{Empirical Application}). 
The forecasting models we consider present the current state-of-the-art, including DCC--GARCH models of \citet{Eng02} and $t$--GAS models inspired by \citet{CKL13}.
Using our proposed methods, one of our findings is that $t$--GAS models and DCC--GARCH models produce equally accurate forecasts for the marginal densities, yet copula forecasts of the latter model are to be preferred.

Once a copula model has been chosen using the methodology proposed in this paper, a natural question is whether the adopted copula provides a sufficiently accurate description of the dependence structure. 
This is important because scores only provide information on the \textit{relative} quality of forecasts, therefore only delivering the ``least bad'' choice.
For checking the \textit{absolute} quality of the copula forecasts, the methods of \citet{ZiegelGneiting2014} may be used.
In this sense, our results nicely complement the aforementioned ones.

The remainder of the paper proceeds as follows.
In Section~\ref{sec:notation}, we introduce some basic notation and definitions.
Section~\ref{Main Results} states our main theoretical results and Section~\ref{Some Two-Step Tests} introduces our two-step tests.
Section~\ref{Simulations} contains the Monte Carlo simulations and Section~\ref{Empirical Application} presents our empirical application.
The final Section~\ref{Conclusion} concludes with a discussion of our results.
To promote flow in the main text, the proofs of all technical results are relegated to the Appendix.

\section{Notation and Definitions}\label{sec:notation}

\subsection{Copulas}

We assume that all random variables in the following are defined on a non-atomic probability space $(\Omega,\mathcal{A},\Pr)$.
Suppose that $\mY=(Y_1,\ldots,Y_d)^\prime$ is the random vector, whose dependence structure is to be modeled.
We denote its joint cumulative distribution function (cdf) by $F$ and the marginal cdfs by $F_i$ ($i=1,\ldots,d$).
Then, by \citeauthor{Skl59}'s \citeyearpar{Skl59} theorem \citep[see also][Theorem~7.3]{MFE15} there exists a cdf $C\colon[0,1]^d\rightarrow[0,1]$ with uniform marginals, called a \textit{copula}, such that
\begin{equation}\label{eq:copula}
	F(\vy) = C\big(F_1(y_1),\ldots, F_d(y_d)\big)\qquad \text{for all }\vy=(y_1,\ldots,y_d)^\prime\in\mathbb{R}^d.
\end{equation}
From this representation it is clear that the copula fully describes the dependence structure of $\mY$, with marginal information absorbed in the $F_i$.
\citet{Skl59} also shows that the copula is unique if all marginal cdfs are continuous.
Vice versa, for any copula $C$ and any marginal distributions $F_1,\ldots,F_d$, the right-hand side of \eqref{eq:copula} defines a cdf on $\R^d$.
Following almost all of the copula literature, we shall assume continuous marginals in this paper, implying that the copula $C$ in \eqref{eq:copula} is unique.

In the sequel, we denote the set of all $d$-dimensional copulas by $\C$, and the set of all $d$-dimensional cdfs with continuous marginals by $\F(\R^d)$.
In particular, $\F(\R)$ denotes the set of all continuous cdfs on $\R$.
The subset $\F([0,1]^d)\subset\F(\R^d)$ contains all cdfs with support on $[0,1]^d$. 
The set $\F$ will always refer to a subset of $\F(\R^d)$.
Using this notation, Sklar's theorem implies the existence of the mapping 
\[
\Cop\colon \F(\R^d) \to \C,
\]
which maps a joint cdf $F\in\F(\R^d)$ with continuous marginals to its unique copula $\Cop(F)\in\C$, satisfying \eqref{eq:copula}.
Finally, a \emph{Fr\'echet class} is a subclass of $\F(\R^d)$ with fixed and given marginal distribution structure $\{F_i\}_{i=1,\ldots, d}$. That means, any two members of a Fr\'echet class only differ in terms of their copulas.

\subsection{Scoring Functions and Elicitability}\label{sec:Scoring Functions and Elicitability}

Consider some functional $\T\colon\F\to\A$, mapping any cdf in $\mathcal{F}$ to some \emph{action domain} $\A$.
In the decision-theoretic framework of \citet{Gne11}, forecasts for $T$ are evaluated and ranked in terms of \emph{scoring functions} $S\colon \A\times\R^d\to\R$. 
They assign a forecast $x\in\A$ the (negatively oriented) score $S(x,\vy)\in\R$ if the observation $\vy\in\R^d$ materializes.
Scoring functions are also often called \textit{loss functions} in the econometric literature and are then denoted by an $L$ \citep{GW06,LRV12,CI20}.
The score $S$ is called \textit{strictly $\F$-consistent for $\T$} if the expected score is uniquely minimized by the correctly specified functional, that is, if
\[
\E\big[S(\T(F),\mY)\big]<\E\big[S(x,\mY)\big]\qquad\text{for all }F\in\F, \ \mY\sim F, \text{ and for all } x\in\A, \ x\neq T(F).
\]
Consequently, strictly consistent scoring functions incentivize truthful forecasting.
To see this, suppose that the goal is to elicit the true belief of the forecaster for the functional $T$.
If the forecaster is ``rewarded'' with the negative score $-S(x,\mY)$ for a forecast $x$ and a verifying observation $\mY\sim F$ that materializes, then her expected reward $\E[-S(x,\mY)]$ is maximized by the true report $x=T(F)$.
In this sense, the scoring function $S$ serves as a ``truth serum'' that elicits the true belief.
This example also explains why we say that \emph{$\T$ is elicitable on $\F$}, if a strictly $\F$-consistent scoring function exists for a functional $\T$.

Strict consistency of a scoring function tells us that lower expected scores are to be preferred, because the expected score is minimized by the true report.
In practice, by the law of large numbers, we can approximate the expected score by the sample average of the scores.
Then, the forecast with the lowest average score is the preferred one. 
The following example makes this concrete.

\begin{example}\label{ex:sd}
Suppose that some functional $T$ of the $\R^{d}$-valued $\{\mY_t\}_{t=1,\ldots,n}$ is to be predicted.
The competing forecasts are given by $\{r_{1t}\}_{t=1,\ldots,n}$ and $\{r_{2t}\}_{t=1,\ldots,n}$.
Moreover, assume that $T$ is elicitable with strictly consistent score given by $S$.
(For concreteness, we may think of $T$ as the mean functional, in which case the square loss $S(x,y)=(x-y)^2$ is strictly consistent; see \citet[Theorem~7]{Gne11}.)
Therefore, the predictions are compared via the \textit{average scores} $\frac{1}{n}\sum_{t=1}^{n}S(r_{it},\mY_t)$ ($i=1,2$), and the set of forecasts with the lower score is favored.
To assess whether differences in the average scores are statistically significant, \citet{DM95} suggest to look at the \textit{average score differences}
\[
	\frac{1}{n}\sum_{t=1}^{n}d_t:= \frac{1}{n}\sum_{t=1}^{n}\big[S(r_{1t},\mY_t)-S(r_{2t},\mY_t)\big]
\]
for testing the null of equal predictive ability that $\E[d_t]=0$.
This leads to what are now called Diebold--Mariano tests; variations of which we consider in Section~\ref{Some Two-Step Tests} below.
\end{example}

\begin{rem}\label{rem:misspecification}
Strictly speaking, strict consistency only implies that the expected score is uniquely minimised by the true report. 
The definition itself does not give us any theoretical guarantee how possibly misspecified forecasts are ranked in expectation; see \cite{Pat20}.
Notable exceptions are order-sensitivity \citep{BelliniBignozzi2015, FisslerZiegel2019} and the sensitivity with respect to increasing information sets \citep{HE14}.
That means we should consider strict consistency as a minimal quality criterion in the choice of a scoring function. Of course, once a (strictly consistent) score is fixed, it will then induce a ranking between any forecasts.
\end{rem}

If the functional $\T$ maps to some finite-dimensional action domain $\A\subseteq \R^k$, we speak of a \emph{point forecast}; e.g., when $\T$ is the mean functional or a quantile.
For $d=k=1$, the prime examples for strictly consistent scores are the square loss, $S(x,y) = (x-y)^2$, for the mean, and the absolute loss, $S(x,y)=|x-y|$, for the median.
On the other hand, \emph{probabilistic forecasts} amount to specifying the full predictive distribution. 
In this case, $\T$ is the identity map on $\F$ and the action domain $\A$ corresponds to $\F$ itself. 
Traditionally, strictly consistent scoring functions for probabilistic forecasts are called \emph{strictly proper scores} or \emph{strictly proper scoring rules} \citep{GR07}.
Examples include the log-score $S_{\log}(F,\vy)=-\log f(\vy)$ (with $f(\cdot)$ denoting the density of $F$) and the continuous ranked probability score (CRPS) $S_{\CRPS}(F,\vy)=\E\Vert\mX-\vy\Vert-(1/2)\E\Vert\mX-\tilde{\mX}\Vert$ for independent $\mX\sim F$ and $\tilde{\mX}\sim F$ \citep{GR07}.

As a matter of fact, the copula functional $\Cop\colon \F(\mathbb{R}^d)\to\C$ we study in this paper bridges the realms of point forecasts and probabilistic forecasts; it summarizes the distribution (like point forecasts), yet with an infinite-dimensional action domain (like probabilistic forecasts).

\section{Main Results}\label{Main Results}

\subsection{Copulas Are Not Elicitable}\label{Copulas Are Not Elicitable}

To establish the non-elicitability of the copula functional $\Cop$, we recall that the convex level sets property is a necessary condition for elicitability \citep{Osband1985,Gne11}.

\begin{lem}\label{lem:CxLS}
A functional $\T\colon\F\to\A$ is elicitable on $\F$ only if $\T$ has convex level sets on $\F$: %
For all $F^0, F^1\in\F$ and $\lambda \in (0,1)$ such that 
$F^\lambda = (1-\lambda)F^0 + \lambda F^1 \in\F$ it holds that
\[
T(F^0) = T(F^1) \implies T(F^\lambda) = T(F^0).
\]
\end{lem}
Even though the original proof of Lemma \ref{lem:CxLS} is provided for point forecasts, that is, $\A\subseteq \R^k$, the proof works exactly the same for any kind of action domain, in particular also for $\A=\C$.
We provide the proof in the Appendix for the sake of completeness.

\begin{prop}\label{prop:CxLS of cop}
On any subclass $\F\subseteq \F(\R^d)$ which contains at least one distribution $F^0$ with bounded support along with any translation $F^1 = F^0(\cdot - \vt)$, $\vt\in\R^d$, and at least one mixture distribution $F^{\lambda} = (1-\lambda)F^0 + \lambda F^1$, $\lambda \in (0,1)$,
the copula functional $\Cop$ fails to have convex level sets, thus failing to be elicitable. 
In particular, the copula functional $\Cop$ does not have convex level sets on $\F(\R^d)$ and is not elicitable.
\end{prop}

At first sight, it may appear that the focus on distributions with bounded support in Proposition~\ref{prop:CxLS of cop} is restrictive.
However, the opposite is the case. 
Even if one considers a very small class of distributions that only contains distributions with bounded support (including their translations and one mixture) do copulas fail to be elicitable on that class.
This then trivially implies that copulas are not elicitable on larger classes of distributions (such as $\F(\R^d)$).

Copulas share the property of non-elicitability with many other functionals, such as the variance, the Expected Shortfall and the mode \citep{Gne11,Hei14}.
The practical implication of the non-elicitability is that copula forecasts cannot be sensibly compared on their own.
This is because no score exists for ranking the copula forecasts that satisfies even the minimal quality criterion of being uniquely minimized in expectation by the true report.

\subsection{Copulas Are Elicitable on Fr\'echet Classes}

In contrast to Proposition~\ref{prop:CxLS of cop}, the copula functional has convex level sets on any Fr\'echet class $\F$.
Recall that a Fr\'echet class is defined by the property that any members $F,G\in\F$ may only differ in their dependence structure, while their marginal distributions coincide, i.e., $F_i=G_i$ for all $i=1,\ldots,d$.
In particular, assuming that the copulas of two distributions in $\F$ coincide therefore means the two distributions already coincide. 
Hence, also any convex combination coincides. 
This is a necessary condition for the following elicitability result.

\begin{prop}
\label{prop:elicitability for fixed marginals}
On any Fr\'echet class $\F\subseteq \F(\R^d)$, the copula functional $\Cop$ is elicitable.
In particular, for the given marginal structure $\{F_i\}_{i=1,\ldots,d}$ (common to all members of $\F$) and any strictly proper scoring rule $S$ on $\F$, the map
\begin{equation}
\label{eq:Constr1}
\C\times \R^d \ni (C,\vy) \mapsto S\Big(C\big(F_1(\cdot), \ldots, F_d(\cdot)\big),\vy\Big)
\end{equation}
is strictly $\F$-consistent for $\Cop$. Similarly, for the given marginal structure $\{F_i\}_{i=1,\ldots,d}$ (common to all members of $\F$) and any strictly proper scoring rule $S$ on $\C$, the map
\begin{equation}
\label{eq:Constr2}
\C\times \R^d \ni (C,\vy) \mapsto S\Big(C,\big(F_1(y_1), \ldots, F_d(y_d)\big)^\prime\Big)
\end{equation}
is strictly $\F$-consistent for $\Cop$.
\end{prop}

The score in \eqref{eq:Constr1} directly draws on Sklar's Theorem \eqref{eq:copula} and exploits the strict propriety of $S$ on $\F$.
On the other hand, the score in \eqref{eq:Constr2} exploits the fact that if $\mY$ has continuous marginals $F_1,\ldots, F_d$, the vector of probability transforms $\big(F_1(Y_1), \ldots, F_d(Y_d)\big)'$ is distributed like the copula of $\mY$.
Hence, the first argument of the score can directly use a copula. 
Interestingly, the score in \eqref{eq:Constr1} stays strictly consistent even without the assumption of continuous marginals. And on larger classes with varying marginal forecasts, it constitutes a strictly proper scoring rule; see Proposition~\ref{prop:elicitability}.
In contrast, \eqref{eq:Constr2} hinges on the assumption of continuous marginals and cannot be extended to a strictly proper score on its own when varying the marginal forecasts.
However, it will be a building block for our multi-objective scores of Theorem~\ref{thm:mo copula}.

\begin{example}\label{ex:DM same marg}
Proposition~\ref{prop:elicitability for fixed marginals} essentially prescribes simple Diebold--Mariano tests to formally compare copula forecasts $C_{1t}$ and $C_{2t}$ \textit{when} the marginal forecasts $F_{1t},\ldots,F_{dt}$ of both models coincide and are correctly specified.
These would then be based on the average score difference
\[
	\frac{1}{n}\sum_{t=1}^{n}d_t=\frac{1}{n}\sum_{t=1}^{n}\bigg[S\Big(C_{1t}\big(F_{1t}(\cdot),\ldots,F_{dt}(\cdot)\big),\mY_t\Big)-S\Big(C_{2t}\big(F_{1t}(\cdot),\ldots,F_{dt}(\cdot)\big),\mY_t\Big)\bigg]
\]
with a score $S$ from \eqref{eq:Constr1}.
(Tests based on scores from \eqref{eq:Constr2} may be constructed similarly.)
Once the forecasts $F_{1t},\ldots,F_{dt}$ of the marginals are misspecified, however, there is no theoretical guarantee that the score is uniquely minimized by the true copula.
Therefore, in practice, one needs to ensure that the marginal forecasts are (approximately) correctly specified.
\end{example}

\begin{example}\label{ex:log score}
For the fixed marginal structure $\{F_i\}_{i=1,\ldots,d}$ and $S$ the log-score $S_{\log}(F,\vy) = -\log f(\vy)$, the score in  \eqref{eq:Constr1} yields 
\begin{equation}
\label{eq:KLIC}
S_{\log}\Big(C\big(F_1(\cdot), \ldots, F_d(\cdot)\big),\vy\Big)
= -\log\Big(c\big(F_1(y_1),\ldots, F_{d}(y_d)\big)\Big) - \sum_{i=1}^{d}\log\big(f_i(y_i)\big)\,,
\end{equation}
where $c$ is the copula density of the predictive copula $C$ and $f_1, \ldots, f_d$ are the densities of the marginal distributions $F_1,\ldots, F_d$. 
On the other hand, the construction principle from \eqref{eq:Constr2} solely yields 
\[
S_{\log}\big(C,F_1(y_1), \ldots, F_d(y_d)\big)
= -\log\Big(c\big(F_1(y_1),\ldots, F_{d}(y_d)\big)\Big)\,.
\]
Obviously, since the marginal structure is fixed, the term $- \sum_{i=1}^d \log f_i(y_i)$ is irrelevant in any test for predictive accuracy.

For their copula model selection procedure, \citet{CF06} and \citet{DPV10} use the KLIC-score.
Interestingly, the KLIC-score corresponds to the negative of \eqref{eq:KLIC}.
However, Proposition \ref{prop:elicitability for fixed marginals} provides a vast array of other strictly consistent scores, e.g.\ scores based on the CRPS, that can be used in model selection.
\end{example}

\subsection{Copulas and Marginals Are Multi-Objective Elicitable}\label{sec:Copulas and Marginals Are Multi-Objective Elicitable}

Proposition~\ref{prop:elicitability for fixed marginals} demonstrates that copula forecasts can be compared via strictly consistent scoring functions \emph{on Fr\'echet classes}, i.e., \textit{if} the forecasts of the marginal distributions structure is fixed and correctly specified. 
When this requirement is not met, it is natural to ask---in the spirit of \citet{Osband1985}---whether the copula is jointly elicitable with some other functional, the marginals being the natural candidates.
Indeed, it is immediate from \eqref{eq:copula} that specifying the copula and the marginal distributions amounts to specifying the complete predictive distribution.
Therefore, the copula and the marginals are jointly elicitable, as we formally state next.

\begin{prop}\label{prop:elicitability}
On any class $\F\subseteq \F(\R^d)$,
$\Cop$ together with the marginal distributions is elicitable. If $S$ is a strictly proper scoring rule on $\F$, then a strictly $\F$-consistent score is given by 
\[
\C\times \big(\F(\R)\big)^d \times \R^d \ni \big(C,\{F_i\}_{i=1,\ldots, d}, \vy\big) \mapsto 
S\Big(C\big(F_1(\cdot), \ldots, F_d(\cdot)\big),\vy\Big)\,.
\]
\end{prop}

Similarly as the pairs (variance, mean) and (Value-at-Risk, Expected Shortfall), the pair (copula, marginals) is elicitable.
Yet, comparing the copula and the marginals jointly as suggested in the above proposition, amounts to a test of the full predictive distribution.
This does not allow us to pinpoint differences in average scores to either the copula or the marginals, completely defeating the purpose of the exercise. 

To tackle this issue, we draw on the concept of multi-objective scoring functions introduced by \citet{FH24}.
In particular, instead of scalar scores (taking values in $\R$), we use two-dimensional scores taking values in $\R^2$.
The use of scalar scores (such as the squared loss) pervades the forecasting literature.
This is due to historical reasons and the fact that scalars can easily be compared using the canonical order $\leq$ on $\R$.
When considering (as we do) scores in $\R^2$, we need to equip $\R^2$ with an order that ultimately allows us to compare forecasts.

A suitable order on $\R^2$ in that context is given by the lexicographic order, $\lex$ \citep{FH24}. 
Recall that on $\R^2$ it holds that $(x_1,x_2)^\prime\lex (y_1,y_2)^\prime$ if $x_1<y_1$ or if ($x_1=y_1$ and $x_2\leq y_2$).
Obviously, the lexicographic order is a total order on $\R^2$, meaning that we can compare any two points of $\R^2$.
This is important because our end goal is to compare predictions for which it is crucial that a conclusive ranking emerges.
For example, while $(3,5)^\prime\lex (5,3)^\prime$, it neither holds that $(3,5)^\prime\geq_{\comp} (5,3)^\prime$ nor $(3,5)^\prime\leq_{\comp} (5,3)^\prime$ with respect to the componentwise order $\leq_{\comp}$ on $\R^2$ (which is not a total order), such that the points cannot be compared.
We shall write $(x_1,x_2)^\prime\slex (y_1,y_2)^\prime$ if $(x_1,x_2)^\prime\lex (y_1,y_2)^\prime$ and $(x_1,x_2)^\prime\neq (y_1,y_2)^\prime$.

\begin{defn}\label{defn:score}
A score $\mS\colon \A\times \R^d\to\R^2$ is \textit{strictly multi-objective $\F$-consistent for a functional $T\colon\F\to\A$ with respect to the lexicographic order $\lex$} if the expected score is uniquely minimized by the correctly specified functional, that is, if
\[
\E\big[\mS(\T(F),\mY)\big]\slex \E\big[\mS(x,\mY)\big]\qquad\text{for all }F\in\F, \ \mY\sim F, \text{ and for all } x\in\A, \ x\neq T(F).
\]
If $T$ has a strictly multi-objective $\F$-consistent score with respect to the lexicographic order, we call it \textit{multi-objective elicitable on $\F$ with respect to the lexicographic order}.
\end{defn}

The next result is one of our main contributions and establishes the multi-objective elicitability of the copula functional together with the marginal distributions with respect to the lexicographic order.
This means that the action domain corresponds to $\A=\C\times \big(\F(\R)\big)^d$.

\begin{thm}\label{thm:mo copula}
On any class $\F\subseteq \F(\R^d)$,
$\Cop$ together with the marginal distributions is multi-objective elicitable with respect to the lexicographic order. 
In particular, if $S_i$, $i=1, \ldots, d$, are strictly proper scoring rules on $\F(\R)$ and $S$ is a strictly proper scoring rule on $\F([0,1]^d)$, then a strictly $\F$-consistent score is given by 
$\mS\colon \C\times \big(\F(\R)\big)^d \times \R^d \to(\R^{2}, \lex)$,
\begin{equation}
\label{eq:S bivariate}
\mS \big(C,\{F_i\}_{i=1,\ldots, d}, \vy\big) := 
\begin{pmatrix}
S_{\marg}\big(\{F_i\}_{i=1,\ldots, d}, \vy\big)\\
S_{\cop}\big(C,\{F_i\}_{i=1,\ldots, d}, \vy\big)\
\end{pmatrix}
:=
\begin{pmatrix}
\sum_{i=1}^d S_i(F_i,y_i)\\
	S\Big(C,\big(F_1(y_1), \ldots, F_d(y_d)\big)^\prime\Big)
\end{pmatrix}\,.
\end{equation}
\end{thm}

Similarly to \textit{univariate} strictly consistent scores, our \textit{bivariate} strictly consistent multi-objective scores also play the role of a ``truth serum'': Suppose that, in a first step, the forecaster receives a ``reward'' of $-S_{\marg}\big(\{F_i\}_{i=1,\ldots, d}, \mY\big)$ if she issues the forecasts $\{F_i\}_{i=1,\ldots, d}$ and $\mY$ materializes and, in a second step, gets $-S_{\cop}\big(C,\{F_i\}_{i=1,\ldots, d}, \mY\big)$ for the copula forecast $C$.
Then, in expectation, the \textit{step-wise} reward-maximizing strategy is to report the true marginals of $\mY$ in the first step and the true copula in the second.

Interestingly, the second component in \eqref{eq:S bivariate}, $S_{\cop}$, coincides with the score in \eqref{eq:Constr2} of Proposition~\ref{prop:elicitability for fixed marginals}.
The reason is that Theorem \ref{thm:mo copula} nests the situation of Proposition~\ref{prop:elicitability for fixed marginals}. 
The latter treats the situation when the marginal distribution structure is given and fixed. Hence, in a comparison with the bivariate scores from \eqref{eq:S bivariate}, the first component $S_{\marg}$ is the same, such that the comparison solely hinges on the second component, $S_{\cop}$.

\begin{rem}\label{rem:limitations}
We would like to discuss the limitations of the multi-objective elicitability result of Theorem~\ref{thm:mo copula}.
The asymmetric nature of the lexicographic order implies that a correctly specified copula forecast outperforms any other copula forecast in expectation when ranked with a multi-objective score from \eqref{eq:S bivariate} \emph{if} it is accompanied by correctly specified marginal forecasts. 
If the marginal forecasts are, however, incorrectly specified, we do not have such a theoretical guarantee. 
Indeed, the strict consistency of $S_{\cop}$ hinges on the fact that the distribution of the probability transforms $\big(F_1(Y_1), \ldots, F_d(Y_d)\big)'$ coincides with the copula of $\mY$ if and only if the marginal distributions are correctly specified.
This limitation is not unique to our situation, but always occurs for strictly consistent scoring functions for functionals containing a non-elicitable component, such as the pair Value-at-Risk and Expected Shortfall. I.e., a misspecification of the Value-at-Risk forecast implies that the correctly specified Expected Shortfall forecast does not outperform any other Expected Shortfall forecast in expectation.
More generally, strictly consistent scoring functions provide only limited theoretical guarantees for the ranking of possibly misspecified forecasts; see Remark~\ref{rem:misspecification}.

To safeguard against this limitation, forecast comparisons using the multi-objective scores from Theorem~\ref{thm:mo copula} can be accompanied by an assessment of \emph{probabilistic calibration}, that is, an assessment of the uniformity of the probability transforms $F_1(Y_1), \ldots, F_d(Y_d)$; see
\cite{GneitingRanjan2013}.
We close this discussion by remarking that in our simulation study of Section~\ref{Simulations}, we deliberately examined the robustness of the consistency of $S_{\cop}$ when the marginal forecasts are misspecified; see setting (iii) in Section~\ref{Simulations}. The consequences of this marginal misspecification are manifested in a loss of power when comparing copula forecasts. 
\end{rem}

\section{Some Two-Step Tests}\label{Some Two-Step Tests}

We now demonstrate how to use the theoretical results of the previous section for formal hypothesis testing.
The $\mathbb{R}^{d}$-valued verifying observations for the copula forecasts are denoted by $\mY_1,\ldots,\mY_n$ and the information set relevant to the forecaster at time $t$ is $\mathcal{F}_{t}$.
Examples include the trivial sigma field $\mathcal{F}_{t}=\{\emptyset,\Omega\}$ (i.e., unconditional ``forecasting'') or the information set generated by past observations $\mathcal{F}_{t}=\sigma(\mY_{t},\mY_{t-1},\ldots,)$.
However, the precise content of $\mathcal{F}_{t}$ does not matter in the following.
Let us denote by $F_{1t}$ and $F_{2t}$ the competing $\mathcal{F}_{t-1}$-measurable distributional forecasts for the cdf of $\mY_t\mid\mathcal{F}_{t-1}$ ($t=1,\ldots,n$).
From these, the marginal forecasts $F_{1it}$ and $F_{2it}$ ($i=1,\ldots,d$), and the copula forecasts $C_{1t}$ and $C_{2t}$ can be obtained.

Since the score is the metric by which to judge forecast quality (also for multi-objective scores), tests comparing predictive accuracy are based on the \emph{bivariate} score differences
\begin{equation}\label{eq:(sd)}
	\vd_t=\begin{pmatrix}d_{m,t}\\ d_{c,t} \end{pmatrix} = \begin{pmatrix}
	S_{\marg}(\{F_{1it}\}_{i=1,\ldots,d},\mY_t) 
	- S_{\marg}(\{F_{2it}\}_{i=1,\ldots,d},\mY_t) \\ 
	S_{\cop}(C_{1t},\{F_{1it}\}_{i=1,\ldots,d}, \mY_t) 
	- S_{\cop}(C_{2t},\{F_{2it}\}_{i=1,\ldots,d},\mY_t) 
	\end{pmatrix}.
\end{equation}
When the marginal forecasts are identical, then $d_{m,t}=0$ such that there can only be differences in the second component $d_{c,t}$.
In this case, one can use the standard \citet{DM95} procedure to test equal predictive ability (i.e., $\E[d_{c,t}]=0$) or superior predictive ability (i.e., $\E[d_{c,t}]\leq0$) in the copula component; see Example~\ref{ex:DM same marg}.
Recall that, as a lower score is preferred, the null $\E[d_{c,t}]\leq0$ corresponds to the situation where the first set of forecasts is superior (or, more precisely, non-inferior) to the second set of forecasts.

Therefore, the case different from the standard testing framework arises when the marginal forecasts are different.
The two null hypotheses we consider in this case are
\begin{align*}
	H_0^{=}&\colon\quad \E[\vd_t]=\vzero,\qquad t=1,\ldots,n,\\
	H_0^{\lex}&\colon\quad \E[d_{m,t}]=0\quad\text{and}\quad \E[d_{c,t}]\leq0,\qquad t=1,\ldots,n.
\end{align*}
The two-sided hypothesis $H_0^{=}$ simply tests equal predictive accuracy, while the ``one and a half-sided'' hypothesis $H_0^{\lex}$ tests whether the marginals are equally accurate and the first set of copula forecasts is non-inferior to the second set.

We stress that, despite the fact that we are interested only in the copula forecasts in this paper, it is essential to also test for equally accurate marginals (i.e., $\E[d_{m,t}]=0$) in both $H_0^{=}$ \textit{and} $H_0^{\lex}$.
To see this, recall that due to the non-elicitability of copulas (Proposition~\ref{prop:CxLS of cop}), copula forecasts cannot generally be compared on their own.
Only when the marginals are equally accurate (a special case of this being identical marginal forecasts), is it possible to sensibly compare copula forecasts.
This is because only when $\E[d_{m,t}]=0$ (such that $\E[S_{\marg}(\{F_{1it}\}_{i=1,\ldots,d}, \mY_t)] = \E[S_{\marg}(\{F_{2it}\}_{i=1,\ldots,d}, \mY_t)]$) does the lexicographic order consider the copula component, such that the ordering of $\E[S_{\cop}(C_{1t},\{F_{1it}\}_{i=1,\ldots,d}, \mY_t) ]$ and $\E[S_{\cop}(C_{2t},\{F_{2it}\}_{i=1,\ldots,d}, \mY_t) ]$ becomes decisive in preferring one forecast over the other.
Therefore, it is essential in $H_0^{=}$ and $H_0^{\lex}$ to also test $\E[d_{m,t}]=0$.

We test both $H_0^{=}$ and $H_0^{\lex}$ via the \emph{average score differences} $\overline{\vd}_n:=(\overline{d}_{m,n}, \overline{d}_{c,n})^\prime:=\frac{1}{n}\sum_{t=1}^{n}\vd_t$, which form the sample counterpart to $\E[\vd_t]$.
Let $c_{1n}$ and $c_{2n}$ denote two critical values, whose precise choice is specified in Theorem~\ref{thm:2Step DM} below.
We are inclined to reject $H_0^{=}$ if in a first step $\sqrt{n}|\overline{d}_{m,n}|>c_{1n}$ or---if the first step did not lead to a rejection---it holds in a second step that $\sqrt{n}|\overline{d}_{c,n}|>c_{2n}$.
Our test of $H_0^{\lex}$ proceeds similarly except that the second step rejects if $\sqrt{n}\overline{d}_{c,n}>c_{2n}$.

Since $\overline{d}_{m,n}$ and $\overline{d}_{c,n}$ are generally correlated, a crucial quantity in computing $c_{1n}$ and $c_{2n}$ is an estimate of the long-run variance $\mOmega_n=\Var(\sqrt{n}\overline{\vd}_n)$ of the average score differences.
We estimate $\mOmega_n$ via
\begin{multline*}
	\widehat{\mOmega}_n=\begin{pmatrix}\widehat{\sigma}_{mm,n} & \widehat{\sigma}_{mc,n} \\ \widehat{\sigma}_{mc,n} & \widehat{\sigma}_{cc,n}\end{pmatrix}=\frac{1}{n}\sum_{t=1}^{n}(\vd_t-\overline{\vd}_n)(\vd_t-\overline{\vd}_n)^\prime \\ 
	+ \frac{1}{n}\sum_{h=1}^{m_n}w_{n,h}\sum_{t=h+1}^{n}\Big[(\vd_t-\overline{\vd}_n)(\vd_{t-h}-\overline{\vd}_n)^\prime + (\vd_{t-h}-\overline{\vd}_n)(\vd_t-\overline{\vd}_n)^\prime\Big],
\end{multline*}
where $m_n$ is an integer sequence of cutoffs and $w_{n,h}$ is a triangular array of scalar weights.
Both $m_n$ and $w_{n,h}$ are specified in more detail in the following assumption.

\begin{assumption}\label{ass:DM}
\begin{enumerate}

	\item[B1:] There is some $\Delta_{\vd}<\infty$ such that $\E|\vd_t^\prime\vd_t|^{r}\leq\Delta_{\vd}$ for all $t\ge1$, where $r>2$.
	
	\item[B2:] $\{\vd_t\}$ is $\alpha$-mixing of size $-2r/(r-2)$ or $\phi$-mixing of size $-r/(r-1)$ for $r$ from B1.
	
		\item[B3:] The sequence of integers $m_n$ satisfies $m_n\rightarrow\infty$ and $m_n=o(n^{1/4})$, as $n\to\infty$.
		
	\item[B4:] There is some $\Delta_{w}<\infty$ such that $|w_{n,h}|\leq\Delta_{w}$ for all $n\in\mathbb{N}$ and $h\in\{1,\ldots,m_n\}$, and $w_{n,h}\rightarrow1$, as $n\to\infty$, for all $h=1,\ldots,m_n$.
	
	\item[B5:] $\mOmega_n\overset{\p}{\longrightarrow}\mOmega$, as $n\to\infty$, where $\mOmega$ is positive definite.
\end{enumerate}
\end{assumption}

The point of Assumption~\ref{ass:DM}, which is similar to the conditions in Theorem~4 of \citet{GW06}, is merely to ensure that a multivariate central limit theorem holds for $\vd_t$ and that $\mOmega$ can be estimated consistently. 
Hence, it can be replaced by any other assumption implying these two properties.
We refer to \citet{GW06} for some extensive discussion of the wide range of (marginal and copula) estimation schemes covered by the mixing assumption B2.

\begin{thm}\label{thm:2Step DM}
Suppose Assumption~\ref{ass:DM} holds and let $\alpha\in(0,1)$ denote the significance level. Then:
\begin{itemize}

	\item[(i)] Two-Step Test of $H_0^{=}$: Let the critical values $c_{1n}$ and $c_{2n}$ be chosen to satisfy
		\begin{equation}\label{eq:cv H0eq}
		\Pr\{|Z_{1n}|>c_{1n}\}+\Pr\{|Z_{1n}|\leq c_{1n}, |Z_{2n}|>c_{2n}\}=\alpha
	\end{equation}
	for $\mZ_n=(Z_{1n}, Z_{2n})^\prime\sim \mathcal{N}(\vzero,\widehat{\mOmega}_n)$.
	If we reject $H_0^{=}$ when either
\begin{enumerate}
	\item[1.] $\sqrt{n}|\overline{d}_{m,n}|>c_{1n}$ or, if that is not the case,
	\item[2.] $\sqrt{n}|\overline{d}_{c,n}|>c_{2n}$, 
\end{enumerate}
then this two-step test has asymptotic size $\alpha$. 

 \item[(ii)] Two-Step Test of $H_0^{\lex}$: Let the critical values $c_{1n}$ and $c_{2n}$ be chosen to satisfy
\begin{equation}
	\Pr\{|Z_{1n}|>c_{1n}\}+\Pr\{|Z_{1n}|\leq c_{1n},\; Z_{2n}\;>c_{2n}\}=\alpha \label{eq:cv H0eq2}
\end{equation}
for $\mZ_n=(Z_{1n}, Z_{2n})^\prime\sim \mathcal{N}(\vzero,\widehat{\mOmega}_n)$.
If we reject $H_0^{\lex}$ when either
\begin{enumerate}
	\item[1.] $\sqrt{n}|\overline{d}_{m,n}|>c_{1n}$ or, if that is not the case,
	\item[2.] $\sqrt{n}\overline{d}_{c,n}>c_{2n}$, 
\end{enumerate}
then this two-step test has asymptotic size $\alpha$. 
\end{itemize}
\end{thm}

\begin{figure}[t!]
	\centering
		\includegraphics[width=\textwidth]{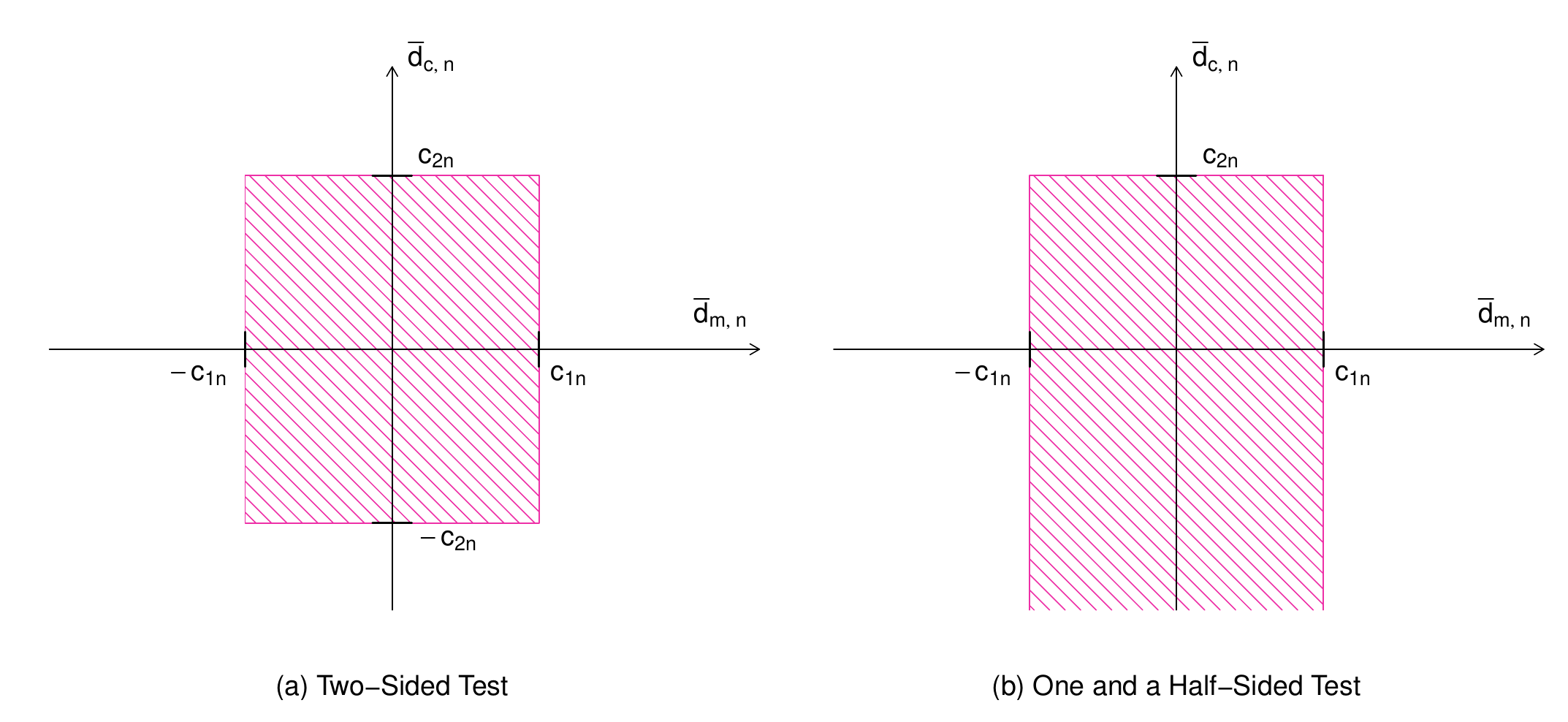}
	\caption{Non-rejection regions of two-step tests for $H_0^{=}$ in (a) and for $H_0^{\lex}$ in (b).}
		\label{fig:NR}
\end{figure}

Theorem~\ref{thm:2Step DM} amounts to a step-wise application of the standard Diebold--Mariano test.
Yet, the critical values are chosen differently, such that the overall (asymptotic) size of our two-step test is preserved.
Figure~\ref{fig:NR} depicts the non-rejection regions of the above two-step tests for $H_0^{=}$ in panel (a) and for $H_0^{\lex}$ in panel (b).
These non-rejection regions depend on the critical values $c_{1n}$ and $c_{2n}$, which are implicitly determined by \eqref{eq:cv H0eq} and \eqref{eq:cv H0eq2}.
In our numerical examples, we choose $c_{1n}$ and $c_{2n}$ such that both probabilities on the left-hand sides of \eqref{eq:cv H0eq} and \eqref{eq:cv H0eq2} are equal to $\alpha/2$, respectively.
Note that the computation of rectangular probabilities for multivariate normal distributions (as required for the computation of $c_{1n}$ and $c_{2n}$) used to be a computational challenge \citep{Gen04}.
Nowadays, this no longer poses a problem using, e.g., the \texttt{R} package \texttt{mvtnorm} \citep{BG09}.

We mention that while we have adopted the standard \citet{DM95} testing framework here, it would also be possible to follow, e.g., the fixed-smoothing approach to predictive ability testing by \citet{CI20}.
We omit details for brevity.

\begin{rem}
\begin{enumerate}
	\item[(i)] It is important not to test $H_0^{=}$ with a standard Wald test, because this would amount to a \emph{joint} test of the marginals and the copula, i.e., a test of the complete predictive distribution without the possibility of attribution.
Since Wald tests are known to possess some optimality properties \citep{Eng84}, our two-step test may trade off some power for a clear attribution of a rejection to either the quality of the marginal or the copula forecasts.

\item[(ii)] It is likewise not advisable to test our hypotheses by employing multiple testing procedures, such as the Bonferroni correction in the simplest case.
The Bonferroni correction would work for $H_0^{=}$ (details for $H_0^{\lex}$ are similar) by choosing critical values $c_{1n}$ and $c_{2n}$ satisfying
\[
	\Pr\{|Z_{1n}|>c_{1n}\}+\Pr\{|Z_{2n}|>c_{2n}\}=\alpha,
\]
where the probabilities are usually assigned equal weight, such that $\Pr\{|Z_{1n}|>c_{1n}\}=\Pr\{|Z_{2n}|>c_{2n}\}=\alpha/2$.
A rejection would then occur when $\sqrt{n}|\overline{d}_{m,n}|>c_{1n}$ or $\sqrt{n}|\overline{d}_{c,n}|>c_{2n}$, such that the Bonferroni test is not carried out in a step-wise manner.
The Bonferroni procedure then has size bounded above by $\alpha$, with actual size typically smaller than $\alpha$.
It is obvious that the critical values implied by \eqref{eq:cv H0eq} can be chosen to be smaller while still keeping size, thus improving power.
Moreover, as for the Wald test, a Bonferroni test cannot be used for properly attributing a rejection to the marginals or the copula. 
\end{enumerate}
\end{rem}

\section{Simulations}\label{Simulations}

As a data-generating process (DGP) we consider a $d$-variate CCC--GARCH model \citep{Bol90}.
Such models have successfully replicated the (co-)movements of returns on speculative assets.
We consider more recent models in this tradition in the empirical application of Section \ref{Empirical Application}.
However, for the present purpose, we prefer a simpler model.
Let $\mathcal{F}_{t-1}=\sigma(\mY_{t-1},\mY_{t-2},\ldots)$ denote the information set consisting of past observables. 
These are generated by the standard CCC--GARCH specification
\[
	\mY_t=(Y_{1t},\ldots,Y_{dt})^\prime=\mSigma_t(\vtheta^\circ)\vvarepsilon_t,\qquad t=1,\ldots,n.
\]
Here, $\mSigma_t(\vtheta^\circ)=\diag(\vsigma_t)$ with $\vsigma_t=(\sigma_{1t},\ldots,\sigma_{dt})^\prime$ is the $\mathcal{F}_{t-1}$-measurable diagonal matrix containing the volatilities of $\mY_t$ (given $\mathcal{\F}_{t-1}$) with true parameter vector $\vtheta^\circ$.
The multiplicative $\R^d$-valued Gaussian noise $\vvarepsilon_t$ is independent of $\mathcal{F}_{t-1}$ and independent, identically distributed (i.i.d.) with mean zero, unit variance and correlation matrix $\mR$ (written $\vvarepsilon_t\overset{\text{i.i.d.}}{\sim}\mathcal{N}(\vzero,\mR)$).

The individual volatilities follow GARCH(1,1) dynamics with identical parameters across $i$, such that $\sigma_{it}^2=\sigma_{it}^2(\vtheta^\circ)=\omega^\circ+\alpha^\circ Y_{i,t-1}^2+\beta^\circ \sigma_{i,t-1}^2$ for $i=1,\ldots,d$ and $\vtheta^\circ=(\omega^\circ, \alpha^\circ, \beta^\circ)^\prime$.
For the parameters, we choose $\omega^\circ=0.001$, $\alpha^\circ=0.1$, $\beta^\circ=0.5$, and we let $\rho=0.5$ be the equicorrelation parameter of the correlation matrix $\mR$.
Hence, we assume that $\mR$ has ones on the main diagonal and $\rho$ everywhere else. 
For the dimension, we set $d=5$.
The sample sizes we consider are $n\in\{150,\ 300\}$. 
Our choices of $d$ and $n$ are inspired by the empirical application in the next section.

By construction $\mY_t\mid\F_{t-1}\sim\mathcal{N}\big(\vzero,\mSigma_t(\vtheta^\circ)\mR\mSigma_t(\vtheta^\circ)\big)$, and we denote the cdf of this normal distribution by $\Phi(\vtheta^\circ,\rho;\mathcal{F}_{t-1})$.
The notation emphasizes the dependence of $\Phi$ on $\F_{t-1}$ here to make clear that this conditional cdf depends on past observables $\mY_{t-1},\mY_{t-2},\ldots$.
From the above, the copula of $\mY_t\mid\F_{t-1}$ is Gaussian with equicorrelation parameter $\rho$, and the conditional marginal distributions of $\mY_t$ are given by $Y_{it}\mid\F_{t-1}\sim\mathcal{N}\big(0,\sigma_{it}^2(\vtheta^\circ)\big)$ for $i=1,\ldots,d$.
Of course, the facts that $\mY_t\mid\F_{t-1}$ has a Gaussian copula with equicorrelation parameter $\rho$ and that $Y_{it}\mid\F_{t-1}\sim\mathcal{N}\big(0,\sigma_{it}^2(\vtheta^\circ)\big)$ ($i=1,\ldots,d$) fully determine the joint distribution of $\mY_t\mid\F_{t-1}$.
Crucially, our construction allows us to separately vary the quality of the marginal forecasts (via $\vtheta^\circ$) and the copula forecasts (via $\rho$).

Specifically, we consider the two distributional ``forecasts'' 
\begin{equation}\label{eq:Fit}
	F_{it}=\Phi\big(\vtheta^\circ\delta^{\marg}_{it},\rho\delta^{\cop}_{it};\mathcal{F}_{t-1}\big),\qquad i=1,2,
\end{equation}
where $\delta^{x}_{it}$ ($i\in\{1,2\}$, $x\in\{\marg, \cop\}$) are mutually independent i.i.d.~draws from a uniform $\mathcal{U}[1-\Delta_i^x,1+\Delta_i^x]$-distribution.
Therefore, the positive $\delta^{x}_{it}$'s may be interpreted as multiplicative noises with $\E[\delta^x_{it}]=1$, which contaminate the true parameters at each point in time.
More precisely, $\delta^{\marg}_{it}$ serves to disturb the forecasts of the marginals and $\delta^{\cop}_{it}$ those of the copula.
The multiplicative disturbances may be interpreted as mimicking estimation error of the true parameters.
However, use of $\delta^{\marg}_{it}$ and $\delta^{\cop}_{it}$ (instead of some actual parameter estimator) allows us to vary the quality of the marginal and copula forecasts independently of each other.
Of course, this comes at the cost of imposing the unrealistic assumption that the true parameters are known.
We discuss this feature of our setup below.
Note that, as in a realistic forecasting setting, the forecasts $F_{it}$ for time $t$ only depend on past information captured by $\mathcal{F}_{t-1}$.

We consider the following five settings for the two competing forecasts $F_{1t}$ and $F_{2t}$, each characterized by a specific choice of $\Delta_1^{\marg}$, $\Delta_2^{\marg}$, $\Delta_1^{\cop}$ and $\Delta_2^{\cop}$:
\begin{enumerate}
	\item[(i)] Size (equally misspecified marginals and copula):\\ 
	$\Delta_1^{\marg}=\Delta_2^{\marg}=0.1$, and $\Delta_1^{\cop}=\Delta_2^{\cop}=0.1$.

	\item[(ii)] Power (equally misspecified marginals, differently misspecified copula): \\
	$\Delta_1^{\marg}=\Delta_2^{\marg}=0.1$, and $\Delta_1^{\cop}=0.5$ and $\Delta_2^{\cop}=0.1$.
	
	\item[(iii)] Power (equally, strongly misspecified marginals, differently misspecified copula):\\
	$\Delta_1^{\marg}=\Delta_2^{\marg}=0.5$, and $\Delta_1^{\cop}=0.5$ and $\Delta_2^{\cop}=0.1$.
	
	\item[(iv)] Power (differently misspecified marginals, equally misspecified copula): \\
	$\Delta_1^{\marg}=0.5$ and $\Delta_2^{\marg}=0.1$, and $\Delta_1^{\cop}=\Delta_2^{\cop}=0.1$.
	
	\item[(v)] Power (differently misspecified marginals, differently misspecified copula): \\
	$\Delta_1^{\marg}=0.5$ and $\Delta_2^{\marg}=0.1$, and $\Delta_1^{\cop}=0.5$ and $\Delta_2^{\cop}=0.1$.
\end{enumerate}
For the size results in (i), both forecasts are equally contaminated (i.e., $\Delta_1^{\marg}=\Delta_2^{\marg}$ and $\Delta_1^{\cop}=\Delta_2^{\cop}$), implying that the null hypotheses $H_{0}^{=}$ and $H_{0}^{\lex}$ hold.
In the settings (ii)--(v), the second set of forecasts $F_{2t}$ is always less (or equally) contaminated than the first (i.e., $\Delta_{2}^x\leq\Delta_1^x$ for $x\in\{\marg,\cop\}$).
Therefore, (ii)--(v) all correspond to cases where the alternative to both $H_{0}^{=}$ and $H_{0}^{\lex}$ is true.
The setup in (iii) is analogous to that in (ii), except that both marginals are more seriously misspecified (but equally so).

We mention that it would be desirable for a more realistic setting to actually estimate the parameters of our CCC--GARCH model using, e.g., the QMLE \citep[Section~11.4]{FZ10}.
However, with a realistic parameter estimator, it would be very difficult to construct forecasts whose quality can be varied for the marginals and the copula separately.
Therefore, we have opted for the above method in \eqref{eq:Fit} of ``contaminating'' the true parameters.

In applying our two-step tests from Theorem~\ref{thm:2Step DM} we use the log-score for $S_i$ ($i=1,\ldots,d$) and $S$ to compute our multi-objective score from Theorem~\ref{thm:mo copula}.
Furthermore, we choose the critical values such that the probabilities on the left-hand side of \eqref{eq:cv H0eq} and \eqref{eq:cv H0eq2} are each equal to $\alpha/2$ for $\alpha=0.05$.
In computing $\widehat{\mOmega}_n$ we follow the recommendation of \citet{DM95} and use $w_{n,h}=0$, such that $\widehat{\mOmega}_n$ equals the sample variance of the score differences.

\begin{table}[t!]
		\centering
		\begin{tabular}{llrrrrrrr}
			\toprule
	$n$	 	& Forecasts		& \multicolumn{3}{c}{$H_0^{=}$}					&&  \multicolumn{3}{c}{$H_0^{\lex}$} 	\\
		\cline{3-5} 					\cline{7-9}\noalign{\vspace{0.5ex}}
				&							&	Marginal & Copula & Joint && Marginal & Copula & Joint\\
			\midrule 
	 150 	& (i)					&	 2.3 &     2.5 &    4.8 &&   2.2 &   2.6 &   4.8	\\
				& (ii)				&	 2.2 &    58.8 &   61.0 &&   2.2 &  70.2 &  72.4	\\
				& (iii)				&	 2.3 &    24.4 &   26.7 &&   2.3 &  34.7 &  37.0	\\
				& (iv)				&	35.1 &     2.4 &   37.5 &&  35.0 &   4.3 &  39.3  \\
				& (v) 				&	34.8 &    27.6 &   62.4 &&  34.8 &  35.9 &  70.7	\\
			\midrule
	300 	& (i)					&	 2.3 &     2.7 &    5.0 &&   2.3 &   2.7 &   5.0  	\\
				& (ii)				&	 2.3 &    88.6 &   90.9 &&   2.3 &  92.9 &  95.2	\\
				& (iii)				&	 2.4 &    49.5 &   51.9 &&   2.5 &  60.7 &  63.2	\\
				& (iv)				&	70.0 &     1.6 &   71.6 &&  70.0 &   2.9 &  72.9	\\
				& (v)				  &	70.7 &    24.0 &   94.7 &&  70.7 &  25.9 &  96.6	\\
			\bottomrule
		\end{tabular}
\caption{Rejection frequencies (in \%) of $H_0^{=}$ and $H_0^{\lex}$ given in column ``Joint''. Rejections due to violations detected in the first (second) step in column ``Marginal'' (``Copula'').}
\label{tab:freq}
\end{table}

We draw the following conclusions from the results in Table~\ref{tab:freq}, which are based on 10,000 replications of the tests of $H_0^{=}$ and $H_{0}^{\lex}$ for each of (i)--(v).
The overall rejection frequencies in the column ``Joint'' indicate that size is close to the nominal level of $\alpha=5\%$ (row (i)) and power is high (rows (ii)--(v)).
The columns ``Marginal'' and ``Copula'' decompose the total number of rejections (in column ``Joint'') into rejections that occurred in the first step (for the marginal component) and the second step (for the copula component), respectively.
By construction of our two-step test, these rejection frequencies should each equal $\alpha/2=2.5\%$ under the null.
Table~\ref{tab:freq} shows that this is roughly the case; see rows labeled (i).

Furthermore, we see that if the difference in the forecasts is only in the marginals (setup (iv)), then the null is rejected mainly in the first step.
Similarly, if differences in predictive ability exist exclusively in the copula component, then our test identifies these mainly in the second step, as it should be; see the results for setups (ii)--(iii).
We conclude that attribution using our tests works well.

It is also noteworthy that for setups (ii) and (iii), where the marginals are equally misspecified, the rejection probability in the first step roughly equals the expected $2.5\%$.
Setting (iii), where the copula forecasts are as different as under (ii), shows that it becomes increasingly more difficult to distinguish between the copula forecasts, when the marginals are more seriously misspecified.
The lesson to be learned is that, while the marginals are not formally tested in the second step, they do have an influence on the power of the copula test.
This is because the copula score differences based on $S_{\cop}$ do implicitly depend on the marginal forecasts; see the definition of $S_{\cop}$ in Theorem~\ref{thm:mo copula} and the discussion of Remark~\ref{rem:limitations}.

We generally observe the highest power when there are differences in predictive ability of both marginals and copulas for forecasts (v).
Moreover, we also see the expected increase in power for larger sample sizes $n$.
Finally, the tests of $H_0^{\lex}$ have higher power than those for $H_0^{=}$, since the deviation from the null is in the direction of the alternative of $H_0^{\lex}$.
This is also as expected for one (and a half)-sided tests.

\section{Empirical Application}\label{Empirical Application}

We illustrate our two-step tests for a broad set of international stock market indices.
Specifically, we consider copula forecasts for the joint daily log-returns $\mY_t=(Y_{1t},\ldots,Y_{dt})^\prime$ on the S\&P~500, DAX, CAC~40, Hang Seng Index and Nikkei~225 (such that $d=5$).
Data on the index returns were downloaded from \href{https://www.wsj.com/market-data/quotes}{www.wsj.com/market-data/quotes}.
The dependence structure of these indices are of importance, e.g., in building internationally diversified portfolios.
Our sample contains 7 years of daily data from 2016 to 2022.
For simplicity, we only keep those observations where returns for all indices are available.

We compare the copula forecasts of seven models.
The first two models are benchmark DCC--GARCH processes, one with normally distributed innovations ($\mathcal{N}$--DCC) and one with multivariate Student's $t$-distributed innovations ($t$--DCC).
These models have been found to provide very accurate variance-covariance matrix forecasts \citep{LRV12,LRV13}. 
The next two models we consider are CCC--GARCH processes, as already used in the simulations in Section~\ref{Simulations}.
Again, we use a normal distribution for the innovations ($\mathcal{N}$--CCC) and a $t$-distribution ($t$--CCC).
As our fifth model, we employ the GO--GARCH of \citet{Van02}, which has Gaussian innovations.
The final two competitors are not from the class of multivariate GARCH models, but from the class of generalized autoregressive score (GAS) models \citep{CKL13,Har13}.
In one case, we assume returns to have a conditional normal distribution ($\mathcal{N}$--GAS), where the marginal scales vary over time (using a GAS update), yet the marginal locations and correlations are time-invariant.\footnote{We have also tried to fit this model with time-varying correlations, yet we only obtained nonsensical parameter estimates.}
In the other case, we assume returns whose conditional distribution is multivariate $t$ ($t$--GAS).
Here, the evolution of the univariate scales and the correlation of the multivariate $t$-distribution follow GAS dynamics with identity scaling matrix.
The degrees of freedom parameter, in contrast, is left constant over time for parsimony.
GAS models have become increasingly popular in time series modeling due to their flexibility and their strong empirical performance in forecasting applications.
The webpage \href{https://www.gasmodel.com} {www.gasmodel.com}, in particular, provides some testimony to the increasing popularity of GAS models.
We refer to \citet{Bea19a} for a more complete description of all the above models as well as their implementation in \texttt{R}.

We use a fixed window of 6 years to estimate the models, for which we use the \texttt{rmgarch} package of \citet{rmgarch} and the \texttt{GAS} package of \citet{ABC19}.
We keep the remaining 1 year (corresponding to $\mY_1,\ldots,\mY_n$ with $n=223$) for comparing the copula forecasts.
Throughout, we denote the distributional forecasts by $F_{1t}$ and $F_{2t}$.
From these, the copula forecasts $C_{1t}$ and $C_{2t}$ are derived as well as the respective copula densities $c_{1t}$ and $c_{2t}$.
Likewise, we obtain the forecasts of the marginals $F_{1it}$ and $F_{2it}$ ($i=1,\ldots,d$) and the appertaining density forecasts $f_{1it}$ and $f_{2it}$, respectively.

\begin{table}[t!]
		\centering
		\begin{tabular}{cccccccc}
		\toprule
					& $\mathcal{N}$--DCC  & $t$--DCC & $\mathcal{N}$--CCC & $t$--CCC & GO--GARCH & $\mathcal{N}$--GAS & $t$--GAS \\
\midrule
$\overline{S}_{\marg,n}$ 	& $-13.76$ & $-13.80$ &  $-13.76$ & $-13.86$ & $-13.81$ &  $-6.12$ & $-13.85$	\\
$\overline{S}_{\cop,n}$		& $ -1.30$ & $ -1.33$ &  $ -1.26$ & $ -1.29$ & $ -1.27$ &  $-1.10$ & $ -1.11$   \\
\bottomrule
		\end{tabular}
\caption{Average scores for the marginals ($\overline{S}_{\marg,n}$) and the copula ($\overline{S}_{\cop,n}$).}
\label{tab:avg score}
\end{table}

To compare forecasts, we use the average score differences as defined in \eqref{eq:(sd)}.
In doing so, we use the multi-objective score $\mS$ from Theorem~\ref{thm:mo copula} (Equation~\eqref{eq:S bivariate}) with the log-scores for $S_i$ and $S$ (see Example~\ref{ex:log score}), such that
\begin{align*}
	S_{\marg}(\{F_{1it}\}_{i=1,\ldots,d},\mY_{t})	&=-\sum_{i=1}^{d}\log\big(f_{1it}(Y_{it})\big),\\
	S_{\cop}(C_{1t},\{F_{1it}\}_{i=1,\ldots,d},\mY_t) &= -\log\Big(c_{1t}\big(F_{11t}(Y_{1t}),\ldots, F_{1dt}(Y_{dt})\big)\Big),
\end{align*}
with similar formulae for the second set of forecasts.
To give an impression of the quality of the forecasts issued by the seven models, we display the average scores for the marginals $\overline{S}_{\marg,n}=\frac{1}{n}\sum_{t=1}^{n}S_{\marg}(\{F_{1it}\}_{i=1,\ldots,d},\mY_{t})$ and the copula $\overline{S}_{\cop,n}=\frac{1}{n}\sum_{t=1}^{n}S_{\cop}(C_{1t},\{F_{1it}\}_{i=1,\ldots,d},\mY_{t})$ in Table~\ref{tab:avg score}.
Recall that, everything else equal, a lower average score is preferable.
The magnitude of the average marginal scores seems to be comparable for all models, except for the $\mathcal{N}$--GAS model that has a much higher score.
For the copula component, the GARCH-type models all perform equally well, with the GAS models lagging equally far behind.

More formally, we now provide the results of our two-step tests, as detailed in Section~\ref{Simulations}.
We first carry out tests of $H_0^{=}$, i.e., the joint hypothesis of equal predictive ability in the marginals and the copula.
Table~\ref{tab:H0=} shows the results for all possible pairwise comparisons.
In that table, an ``M'' indicates a rejection of our two-step test already in the first step that compares the marginal forecasts (such that $\sqrt{n}|\overline{d}_{m,n}|>c_{1n}$ holds in the notation of Theorem~\ref{thm:2Step DM}).
On the other hand, a ``C'' signals a rejection only in the second step analyzing the copula forecasts (i.e., $\sqrt{n}|\overline{d}_{c,n}|>c_{2n}$ but $\sqrt{n}|\overline{d}_{m,n}|\leq c_{1n}$).

Comparing the GAS with the GARCH models in Table~\ref{tab:H0=}, we find significant differences in forecast quality already in the marginals for the $\mathcal{N}$--GAS model, but also in the second step test of the copula forecasts for the $t$--GAS.
Between the CCC and the DCC models, we find non-rejections of the null and rejections of it either due to the marginals or the copulas.
Despite the different predictive ability of the various CCC and DCC models, the differences in forecast quality between the GO--GARCH model and the other GARCH models are always insignificant.

\begin{table}[t!]
		\centering
		\begin{tabular}{cccccccc}
		\toprule
$F_{1t}$ & \multicolumn{7}{c}{$F_{2t}$}\\
		\cmidrule(lr){2-8}
					& $\mathcal{N}$--DCC  & $t$--DCC & $\mathcal{N}$--CCC & $t$--CCC & GO--GARCH & $\mathcal{N}$--GAS & $t$--GAS \\
\midrule
$\mathcal{N}$--DCC 	&  \xmark &	\xmark	& \xmark& \xmark& \xmark& \xmark& \xmark 	\\
$t$--DCC						&  0      &	\xmark 	& \xmark& \xmark& \xmark& \xmark& \xmark    \\
$\mathcal{N}$--CCC	&  C			&	0				& \xmark& \xmark& \xmark& \xmark& \xmark   \\
$t$--CCC  					&  M		  &	M				& M			& \xmark& \xmark& \xmark& \xmark  	\\
GO--GARCH						&  0  		&	0				& 0			&  0		& \xmark& \xmark&	\xmark 	\\
$\mathcal{N}$--GAS  &  M 			&	M				& M			&  M		& M			& \xmark& \xmark	\\
$t$--GAS						&  C		 	&	C				& 0			&  C		& 0			&  M		& \xmark \\
\bottomrule
		\end{tabular}
\caption{\textit{Lower triangular part:} Results of a test of $H_0^{=}$ with indicated $F_{1t}$ and $F_{2t}$.  
Here, a ``0'' indicates a non-rejection, and ``M'' and ``C'' indicate a rejection of the null due to the marginal and copula, respectively.}
\label{tab:H0=}
\end{table}

Strictly speaking, a rejection of $H_0^{=}$ only allows us to conclude that forecasts are of different quality.
Therefore, to assess which model should actually be preferred, we also carry out tests of $H_0^{\lex}$.
We do so for all pairwise comparisons in Table~\ref{tab:H0lex}.
Specifically, in Table~\ref{tab:H0lex} we test $H_0^{\lex}$ with the models in the first column issuing the first set of forecasts $F_{1t}$ and the models in the first row corresponding to the second set of forecasts $F_{2t}$.
In other words, the non-inferiority of the column-listed models is tested with respect to the row-listed models.

While this does not mechanically have to be the case, we find that the entries in the lower triangular part of Table~\ref{tab:H0lex} are identical to those in Table~\ref{tab:H0=}.
However, Table~\ref{tab:H0lex} provides a statistically sound indication of which copula forecasts are to be preferred (at least in those cases, where the marginal forecasts perform equally well).
For instance, while we cannot infer anything about the copula forecasts of the $\mathcal{N}$--GAS relative to the other models (because the marginals are of significantly different predictive accuracy), we reject the non-inferiority of the copula forecasts of the $t$--GAS model compared with all CCC and DCC models.
Therefore, the dependence structure predictions of the CCC/DCC model class significantly outperform those of the $t$--GAS.

Turning to the upper triangular part of Table~\ref{tab:H0lex}, we see that the entries with an ``M'' in the lower triangular part necessarily reappear (in transposed form) in the upper triangular part.
This is because the first-step test of the marginals is two-sided.
However, ``C'' entries may be different for the upper and lower triangular parts, because of the one-sided nature of the copula forecast accuracy test. 
For instance, while the $t$--GAS model is inferior in terms of its copula forecasts to the $\mathcal{N}$--DCC model (lower left entry of the table), the $\mathcal{N}$--DCC model is non-inferior to the $t$--GAS model (upper right entry of the table).

\begin{table}[t!]
		\centering
		\begin{tabular}{cccccccc}
		\toprule
$F_{1t}$ & \multicolumn{7}{c}{$F_{2t}$}\\
		\cmidrule(lr){2-8}
					& $\mathcal{N}$--DCC  & $t$--DCC & $\mathcal{N}$--CCC & $t$--CCC & GO--GARCH & $\mathcal{N}$--GAS & $t$--GAS \\
\midrule
$\mathcal{N}$--DCC 	&  \xmark  	&	 0	 		&   0 		&   M  		&   0 	&  M 		& 0 	\\
$t$--DCC						&     0    	&	\xmark 	&   0 		&   M 		&   0  	&  M  	& 0    \\
$\mathcal{N}$--CCC	&     C   	&	 0	 		& \xmark	&   M 		& 	0 	&  M 		& 0   \\
$t$--CCC  					&     M    	&	 M	 		&   M 		& \xmark  &  	0  	&  M  	& 0  	\\
GO--GARCH						&     0    	&	 0	 		&   0 		&   0   	& \xmark&  M  	&	0 	\\
$\mathcal{N}$--GAS  &     M   	&	 M	 		&   M 		&   M   	&   M 	& \xmark& M 	\\
$t$--GAS						&     C   	&	 C	 		&  	C 		&   C   	&   0 	&  M 		& \xmark \\
\bottomrule
		\end{tabular}
\caption{Results of a test of $H_0^{\lex}$ with indicated $F_{1t}$ and $F_{2t}$. 
Here, a ``0'' indicates a non-rejection, and ``M'' and ``C'' indicate a rejection of the null due to the marginal and copula, respectively.}
\label{tab:H0lex}
\end{table}

Finally, we zoom in on one specific comparison of two models, viz.~the $t$--DCC and the $t$--GAS as two benchmarks in their respective model classes.
Here, Table~\ref{tab:H0=} reveals that there is little difference for the marginals, while the differences in the copula forecasts lead to a rejection of $H_0^{=}$.
In contrast, the hypothesis $H_0^{\lex}$ (with $t$--DCC providing the first set of forecasts, and $t$--GAS the second set) cannot be rejected since the copula forecasts of the DCC--GARCH model are non-inferior to those of the GAS model.
(For the marginals, it is clear that we again do not reject by construction of our two-step test.)

\begin{figure}[t!]
	\centering
		\includegraphics[width=\textwidth]{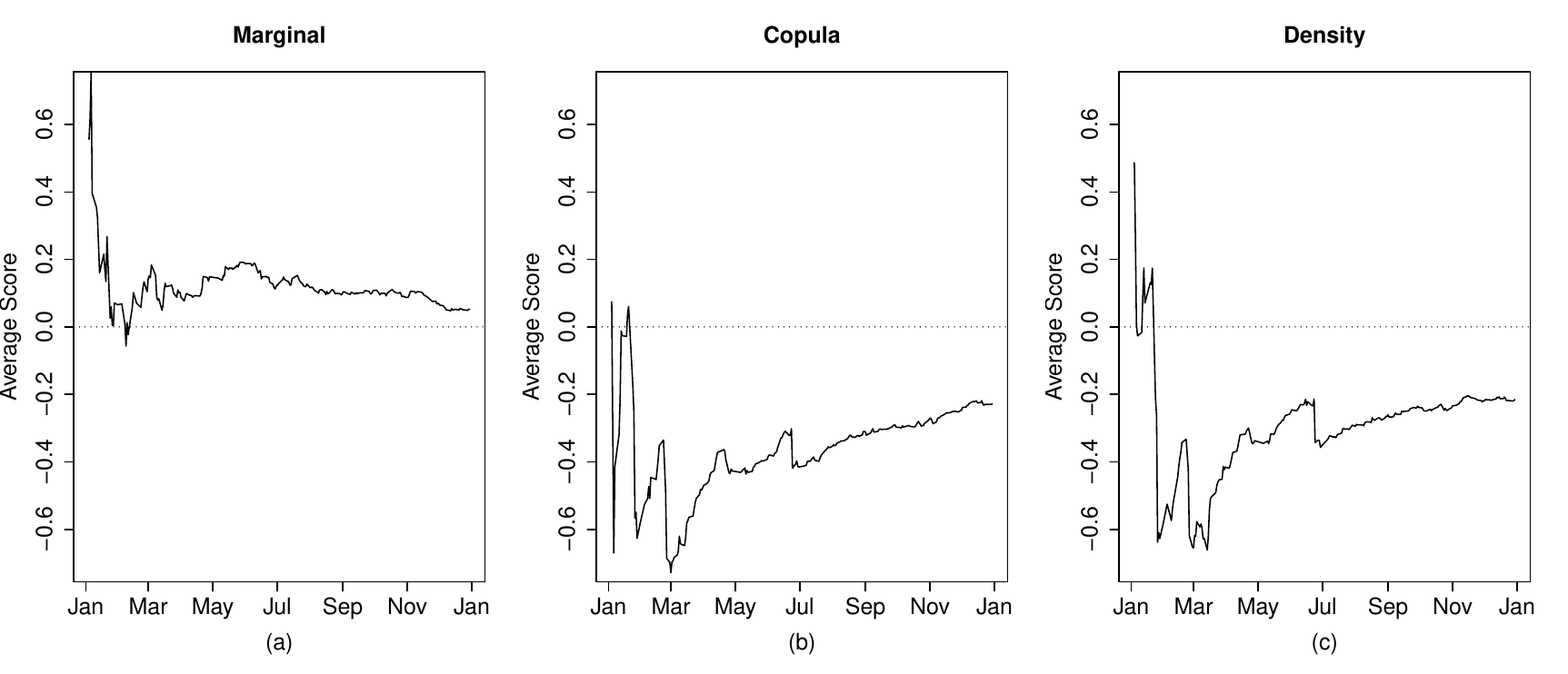}
	\caption{Average score differences over time for (a) marginal, (b) copula and (c) density forecasts during evaluation period in 2022.}
	\label{fig:asd}
\end{figure}

Figure~\ref{fig:asd} illustrates the comparison of the $t$--DCC and the $t$--GAS by plotting the average score differences over time (i.e., $\frac{1}{T}\sum_{t=1}^{T}\vd_t$ for $T=1,\ldots,n$).
For the marginal scores in panel (a), there is evidently little difference.
However, the copula scores in panel (b) are significantly negative---a sign of the superiority of the DCC--GARCH model when it comes to capturing the dependence dynamics over time.

We stress that such an attribution of the rejection to the marginals or the copula would not have been possible by inspecting only the differences in the standard log-scores of the predictive distributions $F_{1t}$ and $F_{2t}$ (panel (c)).
Note that this log-score equals (see Equation~\ref{eq:KLIC})
\begin{equation}\label{eq:ls}
	S_{\log}(F,\vy)=-\log\Big(c\big(F_1(y_1),\ldots, F_{d}(y_d)\big)\Big) - \sum_{i=1}^{d}\log\big(f_i(y_i)\big)\,.
\end{equation}
Therefore, the log-score is the sum of the marginal score $S_{\marg}(\{F_i\}_{i=1,\ldots,d},\vy)$ and the copula score $S_{\cop}(C,\{F_i\}_{i=1,\ldots,d},\vy)$.\footnote{Note that if we had chosen a different scoring rule as building blocks for $S_{\marg}$ and $S_{\cop}$, we would not get such an additive decomposition.}
In particular, adding up the marginal and copula average score differences in panels (a) and (b) yields the average density score differences in panel (c).
While a simple Diebold--Mariano test based on the above log-score in \eqref{eq:ls} likewise reveals that $H_0^{=}$ can be rejected, the reason for the rejection remains unclear.
This contrasts with our two-step test, which clearly attributes the rejection to the quality of the copula forecasts.

\section{Conclusion}\label{Conclusion}

While copulas are widely used in predictive applications in economics and finance, surprisingly little is known about how to carry out a statistically sound forecast comparison.
In this paper, we fill this gap by establishing the following three results.
First, unless the marginal structure is given and fixed, copula forecasts cannot be compared on their own.
Second, when the marginal structure is given and fixed, that is, on Fr\'echet classes, any strictly proper score can validly be used for comparing copula forecasts.
Third, exploiting the existence of multi-objective scores for the copula and the marginals, we demonstrate how to compare 
these joint forecasts while still being able to attribute differences in the forecast accuracy to the quality of the marginal forecasts or the copula forecasts.

Our motivation to introduce multi-objective scores is quite different from that of \citet{FH24}.
Being interested in the comparison of systemic risk forecasts, \citet{FH24} need to overcome the non-elicitability of \citeauthor{AB16}'s \citeyearpar{AB16} conditional Value-at-Risk (CoVaR)---a very popular systemic risk measure.
Even the strategy of combining CoVaR with VaR as an auxiliary functional does not yield elicitability. This motivates \citet{FH24} to introduce multi-objective elicitability---a property that turns out to hold for the pair (CoVaR, VaR).
In our case, the situation is quite different.
The functional of interest, i.e., the copula, \emph{is} jointly elicitable with the marginals (unlike the pair (CoVaR, VaR)).
However, this property does not allow us to pinpoint performance differences specifically to the copula models.
Therefore, the main advantage of multi-objective elicitability (vis-\`{a}-vis elicitability) in our context is that it allows for attributing differences in performance to \emph{either} the marginals \emph{or} the copula.

The non-elicitability result of copulas presented in this paper and the multi-objective elicitability together with the marginal distributions also has consequences for estimation.
It demonstrates why it is generally necessary to first estimate the marginals $\hat F_1, \ldots, \hat F_d$ and then to infer the copula from the pseudo-sample from the copula, $\big\{\big(\hat F_1(Y_{1t}), \ldots, \hat F_d(Y_{dt})\big)'\big\}_{t=1, \ldots, n}$. 
The second step can be done with empirical score minimization of the score in \eqref{eq:Constr2}; see e.g., Chapter 7.5 in \cite{MFE15}.
This two-step procedure amounts to empirical score minimization of the multi-objective scores from Theorem~\ref{thm:mo copula} with respect to the lexicographic order.

In light of the results of this paper and those in \citet{FH24}, we expect there to be many more areas where the concept of multi-objective elicitability may be fruitfully applied.
For instance, in a risk-management context, the pair consisting of VaR and Expected Shortfall (ES) is elicitable and therefore also multi-objective elicitable. 
While forecasts for the pair (VaR, ES) may be compared drawing on its elicitability, it is only the pair's multi-objective elicitability that allows for performance attribution to either the VaR or the ES component.
Explorations such as these are left for future research.

\onehalfspacing

\appendix
\appendixpage

\renewcommand{\theequation}{A.\arabic{equation}}	
\setcounter{equation}{0}

\begin{proof}[{\textbf{Proof of Lemma~\ref{lem:CxLS}:}}]
Suppose that $\T\colon\F\to\A$ is elicitable with a strictly $\F$-consistent score $S$. 
Let $F^0, F^1\in\F$, $\lambda \in (0,1)$ such that 
$F^\lambda = (1-\lambda)F^0 + \lambda F^1 \in\F$ and suppose that $t:=T(F^0) = T(F^1)$ and $x\in\A$, $x\neq t$.
Then
\begin{align*}
\E_{\mY\sim F^{\lambda}}\big[S(t,\mY)\big]
&= (1-\lambda) \E_{\mY\sim F^0}\big[S(t,\mY)\big]
+ \lambda\E_{\mY\sim F^1}\big[S(t,\mY)\big] \\
&< (1-\lambda) \E_{\mY\sim F^0}\big[S(x,\mY)\big]
+ \lambda\E_{\mY\sim F^1}\big[S(x,\mY)\big] 
= \E_{\mY\sim F^{\lambda}}\big[S(x,\mY)\big]\,.
\end{align*}
Invoking the strict $\F$-consistency of $S$, we can deduce that $t = T(F^{\lambda})$.
\end{proof}

\begin{proof}[{\textbf{Proof of Proposition~\ref{prop:CxLS of cop}:}}]
Without loss of generality, we give the proof only for dimension $d=2$.

\textbf{Step 1:}
Let $F^0\in\F$ be a cdf with bounded support, contained in the rectangle $[a,b]\times [c,d]$, $a,b,c,d\in\R$, $a<b, c<d$. 
Let $C$ be the copula of $F^0$. 
Let $F^1$ be a translation of $F^0$ with support contained in the rectangle $[a+t,b+t]\times [c+t,d+t]$, where $t>0$ is such that $a+t>b$ and $d+t>c$.
In other words, the supports of $F^0$ and $F^1$ are disjoint and the support of $F^1$ is on the upper right of the support of $F^0$.
Since copulas are invariant with respect to strictly increasing transformations of each component of a random vector, $F^1$ has the same copula $C$ as $F^0$. 

\textbf{Step 2:}
We first extend the domain of the copula $C$ from $[0,1]^2$ to $\R^2$ by setting $C(x_1,x_2):= C\big(\min(\max(x_1,0),1),\min(\max(x_2,0),1)\big)$ for $x_1,x_2\in\R$. 
A straightforward computation yields that for any $\lambda \in(0,1)$, the copula of the convex combination $F^{\lambda} = (1-\lambda)F^0 + \lambda F^1$ is 
\begin{equation}
\label{eq:C}
C^{\lambda}(u_1,u_2) = 
\frac12\Big[C(2u_1,2u_2) + C\big(2(u_1-\tfrac12), 2(u_2-\tfrac12)\big)\Big]\,,\quad u_1,u_2\in[0,1].
\end{equation}
The left (middle) panel of Figure \ref{fig:Proof} illustrates random samples from $C^\lambda$ if $C$ is the independence (countermonotonicity) copula.

	\begin{figure}
	\centering
		\includegraphics[width=\textwidth]{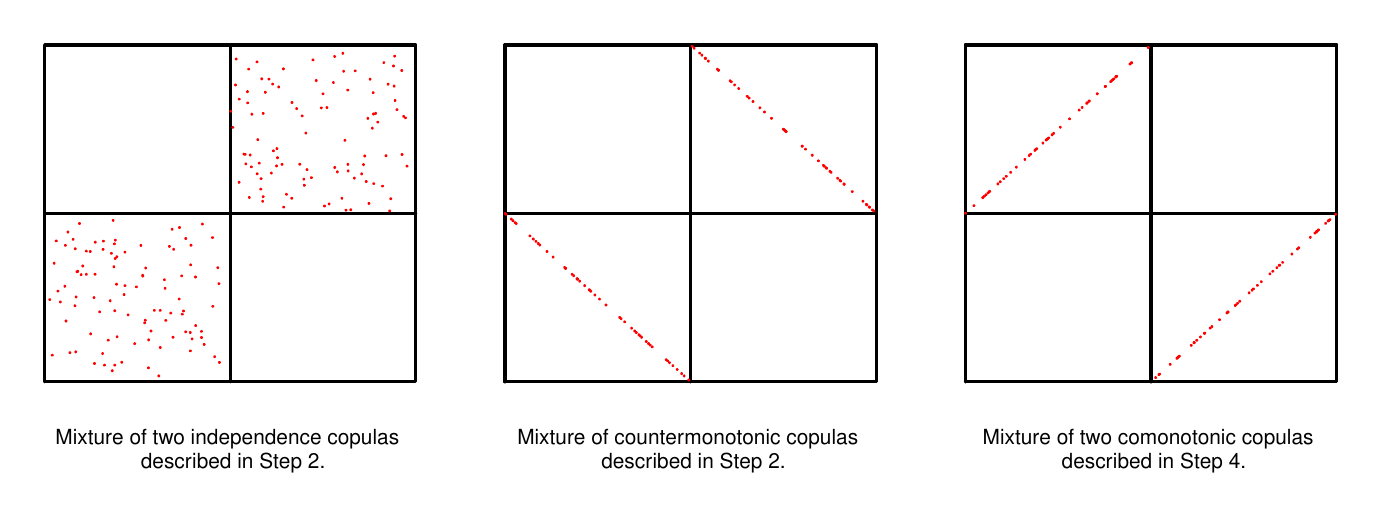}
	\caption{Illustration of random samples from copula mixtures.}
	\label{fig:Proof}
\end{figure}

\textbf{Step 3:}
The copula $C^\lambda$ in \eqref{eq:C} is different from $C$ unless $C$ is the comonotonicity copula (see also Figure \ref{fig:Proof}). 
Indeed, by the construction of $C^\lambda$ in \eqref{eq:C}, $C^{\lambda}$ assigns mass 0 to the rectangles $(0,\tfrac12]\times(\tfrac12,1]$ and $(\tfrac12,1]\times(0,\tfrac12]$, that is, to the upper left and lower right quarters of the unit rectangle. If $C=C^{\lambda}$, then also $C$ must assign mass 0 to $(0,\tfrac12]\times(\tfrac12,1]$ and $(\tfrac12,1]\times(0,\tfrac12]$.
Then again by the identity \eqref{eq:C}, $C^\lambda$ assigns mass 0 to $(0,\tfrac14]\times(\tfrac14,\tfrac12]$, $(\tfrac14,\tfrac12]\times (0,\tfrac14]$, $(\tfrac12, \tfrac34]\times (\tfrac34,1]$ and $(\tfrac34,1]\times (\tfrac12,\tfrac34]$.
Repeating this argument arbitrarily often yields that the assumption $C^\lambda = C$ implies that $C$ assigns mass only to the main diagonal $\{(u,u)^\prime\mid u\in[0,1]\}$. Hence, $C$ must be the comonotonicity copula, recalling the fact that the marginals of the copula must be uniformly distributed.

\textbf{Step 4:}
If the copula of $F^0$ is the comonotonicity copula $M$, consider its translation $F^2$ such that the support of $F^0$ and $F^2$ are disjoint, but the support of $F^2$ is on the \emph{lower right} of the support of $F^0$. Again, also $F^2$ has copula $M$. 
Then, any convex combination of $F^0$ and $F^2$ has the copula
\begin{equation*}
M^{\lambda}(u_1,u_2) = 
\frac12\Big[M\big(2u_1,2(u_2-\tfrac12)\big) + M\big(2(u_1-\tfrac12), 2u_2\big)\Big]\,,\quad u_1,u_2\in[0,1].
\end{equation*}
Clearly, $M^\lambda\neq M$, wich is also illustrated in the right panel of Figure \ref{fig:Proof} where we can see a random sample from $M^\lambda$.
\end{proof}

\begin{proof}[{\textbf{Proof of Proposition~\ref{prop:elicitability for fixed marginals}:}}]
Let $\F\subseteq \F(\R^d)$ be some Fr\'echet class with fixed marginal structure $\{F_i\}_{i=1,\ldots, d}$ common to all members of $\F$.
Let $\mY\sim F\in\F$, $C = \Cop(F)$ and let $C'$ be a copula different from $C$. Then, for a strictly proper scoring rule $S$ on $\F$
\[
\E \Big[S\big(C\big(F_1(\cdot),\ldots, F_d(\cdot)\big),\mY\big)\Big]
= \E  \big[S(F,\mY)\big] < \E  \Big[S\big(C'\big(F_1(\cdot),\ldots, F_d(\cdot)\big),\mY\big)\Big]
\]
by the strict propriety of $S$ on $\F$ and the uniqueness of the copula for distributions with continuous marginals. This establishes the strict $\F$-consistency of the score in \eqref{eq:Constr1}.
On the other hand, if $S$ is a strictly proper scoring rule on $\C$, recall that $\big(F_1(Y_1), \ldots, F_d(Y_d)\big)^\prime \sim C$. Therefore, 
\[
\E\Big[S\big(C,\big(F_1(Y_1), \ldots, F_d(Y_d)\big)^\prime\big)\Big]
< \E\Big[S\big(C',\big(F_1(Y_1), \ldots, F_d(Y_d)\big)^\prime\big)\Big]
\]
by the strict propriety of $S$ on $\C$ and the uniqueness of the copula, which establishes the strict $\F$-consistency of the score in \eqref{eq:Constr2}.
\end{proof}

\begin{proof}[{\textbf{Proof of Proposition~\ref{prop:elicitability}:}}]
This proof is analogous to the proof of the first part of Proposition~\ref{prop:elicitability for fixed marginals}.
The only case left to consider is when one of the marginals is misspecified. Then, by Sklar's theorem \eqref{eq:copula}, the resulting distributional forecast is different from the correctly specified forecast, which yields the claim, invoking the strict propriety of $S$.
\end{proof}

\begin{proof}[{\textbf{Proof of Theorem~\ref{thm:mo copula}:}}]
Let $F\in\F$ and $\mY\sim F$. Let $F_1, \ldots, F_d$ be the true marginal distributions and $C = \Cop(F)$ the true copula. Let $G_1,\ldots, G_d$ be alternative marginal forecasts and $C'$ and alternative copula forecast.
If there is some $j\in\{1,\ldots, d\}$ such that $G_j \neq F_j$, then the strict propriety of $S_j$ implies that $\E\big[S_j(F_j,Y_j)\big]<\E\big[S_j(G_j,Y_j)\big]$. Hence, 
$\sum_{i=1}^d \E\big[S_i(F_i,Y_i)\big] = \E\big[S_{\marg}\big(\{F_i\}_{i=1,\ldots, d}, \mY\big)\big]<\E\big[S_{\marg}\big(\{G_i\}_{i=1,\ldots, d}, \mY\big)\big]$. This implies that 
$$\E\big[\mS \big(C,\{F_i\}_{i=1,\ldots, d}, \mY\big)\big] \slex \E\big[\mS \big(C',\{G_i\}_{i=1,\ldots, d}, \mY\big)\big].$$
On the other hand, if $\{F_i\}_{i=1, \ldots, d} = \{G_i\}_{i=1, \ldots, d}$, then $\big(F_1(Y_1), \ldots, F_d(Y_d)\big)^\prime \sim C$. 
Hence, the strict propriety of $S$ implies that 
\begin{align*}
\E\big[S_{\cop}\big(C,\{F_i\}_{i=1,\ldots, d}, \mY\big)\big]
&= \E\Big[S\Big(C, \big(F_1(Y_1), \ldots, F_d(Y_d)\big)^\prime\Big)\Big] \\
&< \E\Big[S\Big(C', \big(F_1(Y_1), \ldots, F_d(Y_d)\big)^\prime\Big)\Big]\\
&= \E\Big[S_{\cop}\big(C',\{G_i\}_{i=1,\ldots, d}, \mY\big)\Big]\,.
\end{align*}
This implies that 
\[
	\E\big[\mS \big(C,\{F_i\}_{i=1,\ldots, d}, \mY\big)\big] \slex \E\big[\mS \big(C',\{G_i\}_{i=1,\ldots, d}, \mY\big)\big],
\]
which finishes the proof.
\end{proof}

\begin{proof}[{\textbf{Proof of Theorem~\ref{thm:2Step DM}:}}]
Theorem~6.20 in \citet{Whi01} implies that under B1--B4, $\|\widehat{\mOmega}_n-\mOmega_n\|\overset{\p}{\longrightarrow}\vzero$, as $n\to\infty$.
Together with $\mOmega_n\overset{\p}{\longrightarrow}\mOmega$ from B5, this implies that 
\begin{equation}\label{eq:Cons LRV}
	\|\widehat{\mOmega}_n-\mOmega\|\overset{\p}{\longrightarrow}\vzero,\quad \text{as }n\to\infty.
\end{equation}
Next, we show that, as $n\to\infty$,
\begin{equation}\label{eq:std con}
	\sqrt{n}\overline{\vd}_n\overset{d}{\longrightarrow}\mathcal{N}(\vzero, \mOmega),
\end{equation}
where $\mI_{2\times 2}$ denotes the $(2\times 2)$-identity matrix.
The proof is similar to that of Theorem~4 in \citet{GW06}, so we only sketch it here. 
An application of the Cram\'{e}r--Wold device and Theorem~5.20 in \citet{Whi01} ensure that 
$\mOmega_n^{-1/2}\sqrt{n}\overline{\vd}_n\overset{d}{\longrightarrow}\mathcal{N}(\vzero,\mI_{2\times2})$, as $n\to\infty$, under B1 and B2. 
Together with B5, we obtain \eqref{eq:std con}.

Note that size for our two-step procedure is controlled asymptotically under $H_0^{=}$ if
\begin{align*}
	\alpha\underset{(n\to\infty)}{\longleftarrow}&  \Pr_{H_{0}^{=}}\Big\{\big\{\sqrt{n}|\overline{d}_{m,n}|>c_{1n}\big\} \cup \big\{\sqrt{n}|\overline{d}_{c,n}|\leq c_{1n},\ \sqrt{n}|\overline{d}_{c,n}|>c_{2n}\big\} \Big\}\notag\\
	=\,&\Pr_{H_{0}^{=}}\big\{\sqrt{n}|\overline{d}_{m,n}|>c_{1n}\big\} + \Pr_{H_{0}^{=}}\big\{\sqrt{n}|\overline{d}_{m,n}|\leq c_{1n},\ \sqrt{n}|\overline{d}_{c,n}|>c_{2n}\big\}\\
	=:&g_n(c_{1n},c_{2n}).
\end{align*}
To show this, use \eqref{eq:std con} to see for arbitrary (but fixed) $c_{1}$ and $c_{2}$ that, as $n\to\infty$,
\[
	g_n(c_{1},c_{2})\longrightarrow g(c_1,c_2):=\Pr\big\{|Z_1|>c_1\big\} + \Pr\big\{|Z_1|\leq c_1,\ |Z_2|>c_2\big\}
\]
for $\mZ=(Z_1,Z_2)^\prime\sim \mathcal{N}(\vzero,\mOmega)$.
The cdf~of the $\mathcal{N}(\vzero, \mOmega)$-distribution in \eqref{eq:std con} is continuous, such that---by a classical result in real analysis---the convergence $g_n(c_{1},c_{2})\longrightarrow g(c_{1},c_{2})$ is uniform in $c_{1}$ and $c_{2}$.
Moreover, because of \eqref{eq:Cons LRV} it holds for $c_{1n}$ and $c_{2n}$ as specified in the theorem that $c_{1n}\longrightarrow c_1$ and $c_{2n}\longrightarrow c_2$, as $n\to\infty$.
Therefore, $g_n(c_{1n},c_{2n})\longrightarrow g(c_1,c_2)$, as desired.

We omit the proof for $H_0^{\lex}$, which is similar.
\end{proof}

\section*{Supplementary Materials}

The Supplementary Material contains \texttt{R} code that replicates the simulations in Section~\ref{Simulations}.
We also supply \texttt{R} code and data to reproduce the empirical application in Section~\ref{Empirical Application}.

%

%

\section*{Funding}

The work of Yannick Hoga was funded by the Deutsche Forschungsgemeinschaft (DFG, German Research Foundation) through projects 460479886 and 531866675.

\bibliographystyle{apalike}
\bibliography{biblio_Copula}

\end{document}